# A second phase transition and superconductivity in the $\beta$-pyrochlore oxide $KOs_2O_6$

**Zenji Hiroi, Shigeki Yonezawa and Jun-Ichi Yamaura**

Institute for Solid State Physics, University of Tokyo, Kashiwa, Chiba 277-8581, Japan

E-mail: hiroi@issp.u-tokyo.ac.jp

**Abstract**
Another phase transition that is probably of first order is found in the $\beta$-pyrochlore oxide superconductor $KOs_2O_6$ with a superconducting transition temperature $T_c$ of 9.6 K. It takes place at $T_p$ = 7.5 K in the superconducting state in a zero magnetic field. By applying magnetic fields of up to 140 kOe, the $T_c$ gradually decreased to 5.2 K, while $T_p$ changed little, eventually breaking through the $H_{c2}$ line at approximately 65 kOe in the $H$-$T$ diagram. Both the normal-state resistivity and $H_{c2}$ change slightly but significantly across the second phase transition. It is suggested that the transition is associated with the rattling of potassium ions located in an oversized cage of osmium and oxide ions.

## 1. Introduction

The relationship between certain instability and superconductivity has been the subject of intensive experimental and theoretical study. Many exotic superconductors have been found near a critical regime associated with magnetic instability or charge fluctuations [1]. Structural instability in conventional phonon-mediated superconductivity has been a major topic for several decades, because it is believed to be one of the key ingredients for enhancing $T_c$. Indeed, most high-$T_c$ (before the discovery of copper oxide superconductors) materials exhibit some signs of structural instability. In the family of transition metal dichalcogenides, for example, 2H-$NbSe_2$ possesses the lowest charge-density wave transition temperature (33.5 K) and the highest $T_c$ (7.2 K) [2]. The influence of lattice instability associated with a cubic-to-tetragonal transformation on superconductivity has been studied in A-15 compounds [3]. The typical $V_3Si$ compound shows a structural transition at $T_m$ = 21 K and superconductivity below $T_c$ = 17 K. It has been suggested that phonon softening, which accompanies structural instabilities, leads to enhanced $T_c$ values in A-15 compounds [3]. The role of low-energy phonons has also been discussed for Chevrel phase superconductors [4, 5]. The first pyrochlore oxide superconductor, $Cd_2Re_2O_7$ with $T_c$=1.0 K [6], exhibits structural instability that results in two successive structural transitions, from cubic to tetragonal at 200 K and to another tetragonal structure at 120 K [7]. To understand the mechanism of superconductivity, it is crucial to investigate the nature of such an additional transition, because it must reflect relevant instability inherent to the compound.

It is naively expected that geometrical frustration on triangle-based lattices can provide another route leading to novel superconductivity, because it may suppress long-range order that possibly competes with superconducting order. The pyrochlore lattice comprised of corner-sharing tetrahedra is known to be highly frustrated for a localized electron or spin systems with antiferromagnetic nearest-neighbor interactions. Some metallic spinel or pyrochlore oxides which contain the pyrochlore lattice occupied by transition metal ions have been studied extensively in order to elucidate the role of frustration on their physical properties. Recently, another family of pyrochlore oxide superconductors was found, the $\beta$-pyrochlore osmium oxides $AOs_2O_6$ (where $A$ is Cs [8], Rb [9-11] and K [12]) with $T_c$ values of 3.3 K, 6.3 K and 9.6 K, respectively. These have received much attention, because the superconductivity occurs in a frustrated pyrochlore lattice of Os atoms, figure 1. Moreover, the question as to why the $T_c$ is enhanced so much compared with $Cd_2Re_2O_7$ and varies systematically with the $A$ elements is intriguing. Experimental efforts to elucidate the nature of the superconductivity are in progress [13-21]. One puzzling observation is that a few phenomenological properties seem to be quite different among the three members, in spite of the fact that the fundamental electronic structure and the basic pairing mechanism for the superconductivity must be virtually common to all of them. In particular, $KOs_2O_6$, with the highest $T_c$ value, is distinguished from the other compounds, exhibiting various unconventional features. For example, NMR experiments by Arai *et al* showed a tiny coherence peak in the relaxation rate divided by temperature [$1/(T_1T)$] below $T_c$ for $RbOs_2O_6$, while no peaks were found for $KOs_2O_6$ [22]. Moreover, $1/(T_1T)$ is strongly enhanced and temperature-dependent for $KOs_2O_6$ compared with $RbOs_2O_6$, indicating an unusual relaxation process involved in the former [22]. On the other hand, the jump in specific heat at $T_c$ is very large for $KOs_2O_6$, $\Delta C/T_c$ = 185.4 mJ $K^{-2}$ $mol^{-1}$ [23] and 204 mJ $K^{-2}$ $mol^{-1}$ [19], compared with ~40 mJ $K^{-2}$ $mol^{-1}$ [24] and 34 mJ $K^{-2}$ $mol^{-1}$ [11] for $RbOs_2O_6$ and ~26 mJ $K^{-2}$ $mol^{-1}$ for $CsOs_2O_6$ [24].

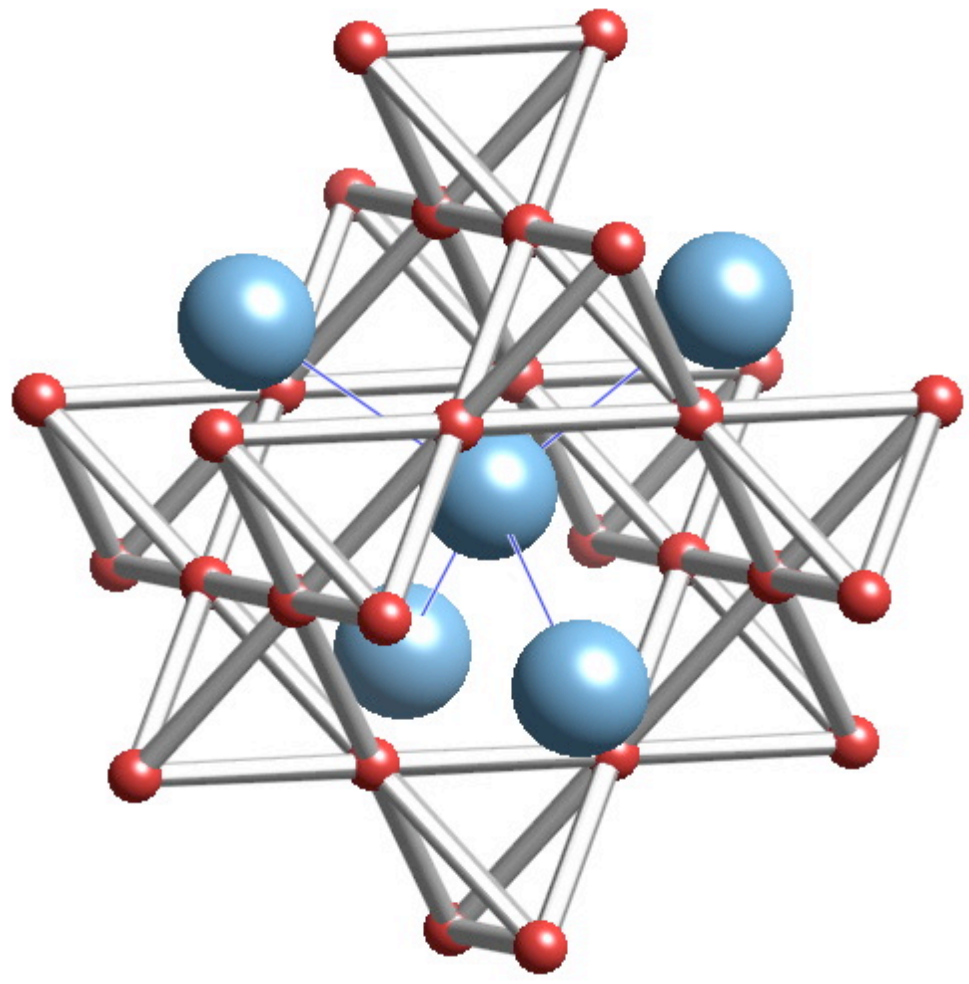

**Figure 1**. (Color online) Crystal structure of $\beta$-$KOs_2O_6$. Os atoms (small ball) form the pyrochlore lattice, and K atoms (big ball) located in the relatively large void of the pyrochlore lattice form a hypothetical diamond lattice.

The temperature dependence of resistivity also discriminates $KOs_2O_6$ from the other compounds: it commonly exhibits an anomalous concave-downward curvature at high temperature, followed by $T^2$ behavior at low temperatures below 30 K and 45 K for the Rb and Cs compounds, respectively [8, 9]. In contrast, the concave-downward resistivity survives even down to $T_c$ for $KOs_2O_6$ [23]. This concave-downward resistivity may not be understood in the framework of either conventional electron-phonon or electron-electron scattering mechanisms, but is probably associated with the rattling phenomenon for the alkaline metal ions [25].

A possible source of structural instability in the $\beta$-pyrochlore oxides is the rattling of alkaline metal ions [24, 26, 27]. Rattling phenomena are generally expected for a relatively small atom located in an oversized atomic cage and have been studied extensively for clathrate and filled skutterudite compounds [28, 29]. A similar situation is realized in the $\beta$-pyrochlore oxides, particularly in $KOs_2O_6$, which has the smallest rattler [26]. It is interesting to investigate how this unusual vibration mode affects superconductivity, because it is an almost localized, incoherent motion in an anharmonic potential and thus is completely different from ordinary low-energy phonons, such as those studied in previous superconductors [3].

The most intriguing finding for $KOs_2O_6$ is the second anomaly in specific heat [19, 23], which has not been found for the other members [24]. A broad hump centered at $T_p$ = 7.5 K, well below $T_c$, was observed in previous specific heat measurements using two blocks of single crystals of $KOs_2O_6$ [23]. It was also reported that $T_p$ changes little under magnetic fields, suggesting that the second anomaly was not directly associated with the superconductivity. However, there was a fatal error in this previous report: the magnitude of the magnetic fields was incorrect and should be reduced by a factor of 0.58 [30]. Thus, the maximum field applied in the experiments was in fact only 80 kOe, not 140 kOe as reported in the paper. In this paper, we report a new set of data covering a wide range of magnetic fields up to 140 kOe using a large, high-quality single crystal of 1 mm in size rather than a block of tiny crystals. Because of the wider range of data and, more importantly, the high quality of the crystal, it was found that the broad hump at $T_p$ for the specific heat became a very sharp peak with thermal hysteresis, indicative of a first-order phase transition. Further evidence of the second transition was obtained by resistivity and magnetization measurements. We demonstrate that the second transition definitely affects both the normal and superconducting properties of $KOs_2O_6$, indicating that there is an interesting relationship between the two transitions.

## 2. Experimental

A single crystal was grown as reported previously [23]. It possesses a truncated octahedral shape with large (111) facets and is approximately 1.0 x 0.7 x 0.3 $mm^3$ in size and 1.302 mg in weight. The electrical resistivity and specific heat were measured in magnetic fields of up to 140 kOe in a Quantum Design physical property measurement system (PPMS). The fields were calibrated by measuring the magnetization of a standard Pd specimen and the voltage of a Hall device (F.W. BELL, BHA-921). Resistivity measurements were carried out by the four-probe method with a current flow along the [1-10] direction and a magnetic field along the [111] direction of the cubic crystal structure. All the measurements were carried out at a current density of 1.5 A $cm^{-2}$. Specific heat measurements were performed by the heat-relaxation method with magnetic fields of up to 140 kOe along the [111] direction, as reported previously [24]. Magnetization was measured in magnetic fields of up to 70 kOe in a Quantum Design magnetic property measurement system and at up to 140 kOe in a PPMS. The magnetic fields were applied approximately along the [1-10] direction.

## 3. Results and Discussion

Figure 2 shows the specific heat of the $KOs_2O_6$ crystal measured on cooling at zero field and in selected magnetic fields of up to 140 kOe. The superconducting transition at zero field takes place with a large jump of $\Delta C/T_c$=194.6 mJ $K^{-2}$ $mol^{-1}$, which is close the previous data [19, 23]. The mean-field $T_c$, defined as the midpoint of the jump, is 9.60 K and the transition width (10-90%) is 0.14 K. $T_c$ gradually decreased with increasing field to reach 5.2 K at 140 kOe. The field dependence of $T_c$ is plotted in the $H$-$T$ phase diagram in figure 6, which should give the $H_{c2}(T)$ curve. It is not easy to estimate the lattice part from the data and extract the electronic contribution, because there is a large contribution from peculiar low-energy phonons coming from the rattling of the K ions, as pointed out previously [23]. Detailed analysis of the phonons and evaluation of the electronic contribution will be reported elsewhere.

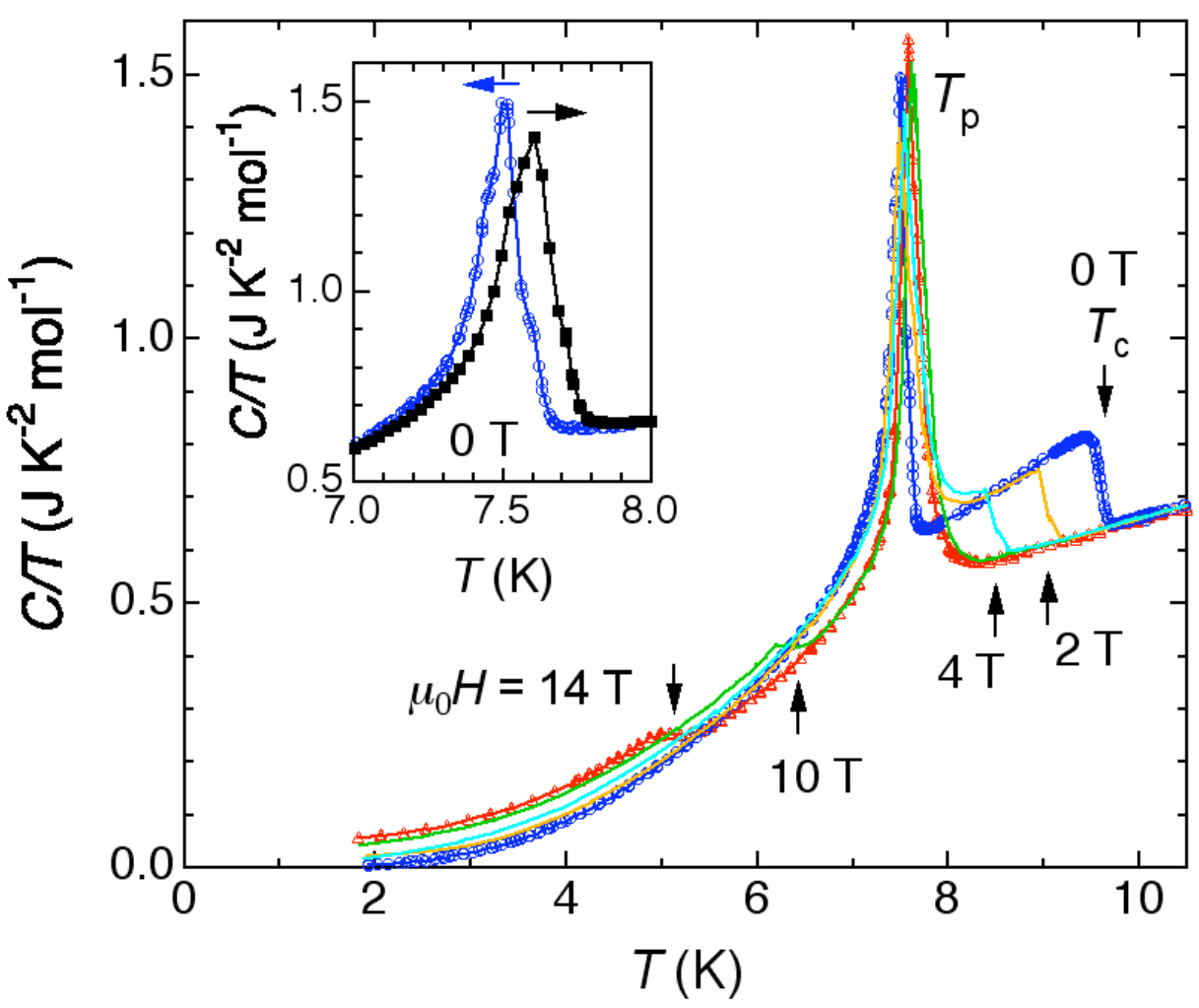

**Figure 2.** (Color online) Specific heat divided by temperature ($C/T$) measured on cooling under a magnetic flux density of 0 T (circles) and 14 T (triangles) applied along the [111] direction. Selected sets of data taken at intermediate fields (2, 4 and 10 T) are also shown. Arrows indicate $T_c$, defined at the midpoint of the jump for each field. $T_p$ is the peak temperature for the second phase transition. The inset shows an enlargement around the second transition. There is thermal hysteresis between the two sets of data measured on cooling (circles) and on heating (squares).

The most striking feature of figure 2 is the appearance of a huge peak at $T_p$ below $T_c$ at zero field, which must correspond to the second anomaly reported previously [19, 23]. Based on an approximate background, the peak height is 1 J $K^{-2}$ $mol^{-1}$, approximately 2.5-fold greater than that for previous data. The inset shows that there is a thermal hysteresis between the two sets of data measured on cooling and heating: the peak appears at ~7.5 K on cooling and at ~7.6 K on heating. Such hysteresis is characteristic of a first-order phase transition. In addition, there are shoulders on either side of the peak. This may reflect a distribution of the critical temperature, possibly arising from inhomogeneity still present in this high-quality crystal. It is expected that the peak would be further enhanced as the sample quality improves. This suggests that the associated phase transition is essentially of first order. We cannot conclude this from the present specific heat measurements using the relaxation method, because the latent heat that would be expected for a first-order transition cannot be measured. The peak slightly shifts to higher temperature by only approximately 0.1 K at 140 kOe. The entropy change associated with the peak is estimated to be 360 mJ $K^{-1}$ $mol^{-1}$.

Figure 3 shows a set of low-temperature resistivity data measured on cooling at various magnetic fields. The zero-resistivity transition temperature $T_{c0}$ is 9.60(1) K at zero field, which is in good agreement with the $T_c$ from specific heat data. The field dependence of $T_{c0}$ is shown in figure 6, which should give the irreversibility field $H_{irr}(T)$ curve in the $H$-$T$ diagram. Surprisingly, a sudden drop in resistivity is observed at a temperature nearly equal to $T_p$ at high magnetic fields above 80 kOe, which was not clearly observed in the previous 80-kOe data (erroneously reported as 140 kOe [23]). This decrease always takes place at $T_p$, independent of the magnitude of the magnetic fields. Small thermal hysteresis between heating and cooling curves is observed, as typically shown for 140 kOe in the inset of figure 3a, which provides further evidence of a first-order phase transition.

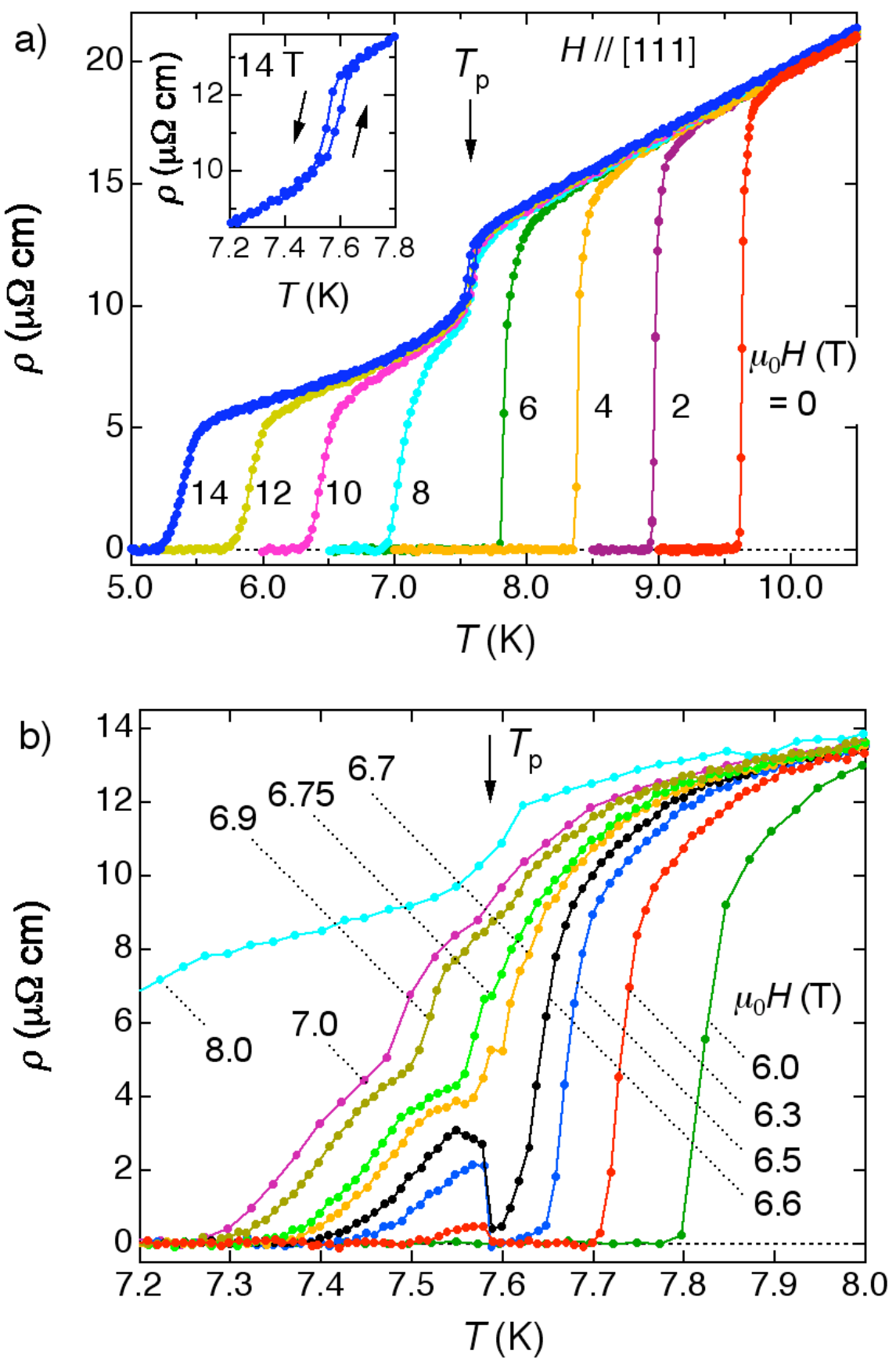

**Figure 3**. (Color online) (a) Evolution of resistivity curves measured on cooling under a magnetic flux density of up to 14 T along the [111] direction. The magnetic flux density was 0, 2, 4, 6, 8, 10, 12 and 14 T, from right to left. The inset shows an enlargement of the data at 14 T near $T_p$ measured on heating after zero-field cooling and then on cooling in the field. (b) Re-entrant superconducting behavior observed in resistivity near $T_p$. The magnetic flux density was 6.0, 6.3, 6.5, 6.6, 6.7, 6.75, 6.9, 7.0 and 8.0 T from right to left.

It should be noted, moreover, that the temperature dependence of the resistivity apparently changes across the transition. As shown in figure 4, it exhibits a concave-downward curvature over a wide temperature range from room temperature to just above $T_p$, as reported previously for a polycrystalline sample [12] or a block of tiny crystals [23]. In contrast, a concave-upward curvature is observed below $T_p$, where the resistivity changes almost linearly as a function of $T^2$. Residual resistivity of 1 μΩ cm was estimated by extrapolating the $T^2$ dependence to $T = 0$. This implies a substantial change in the dominant scattering mechanism of carriers. On the other hand, the drop in resistivity implies either elongation of electron mean free path or an increase in carrier density below $T_p$ arising from a change in the band structure.

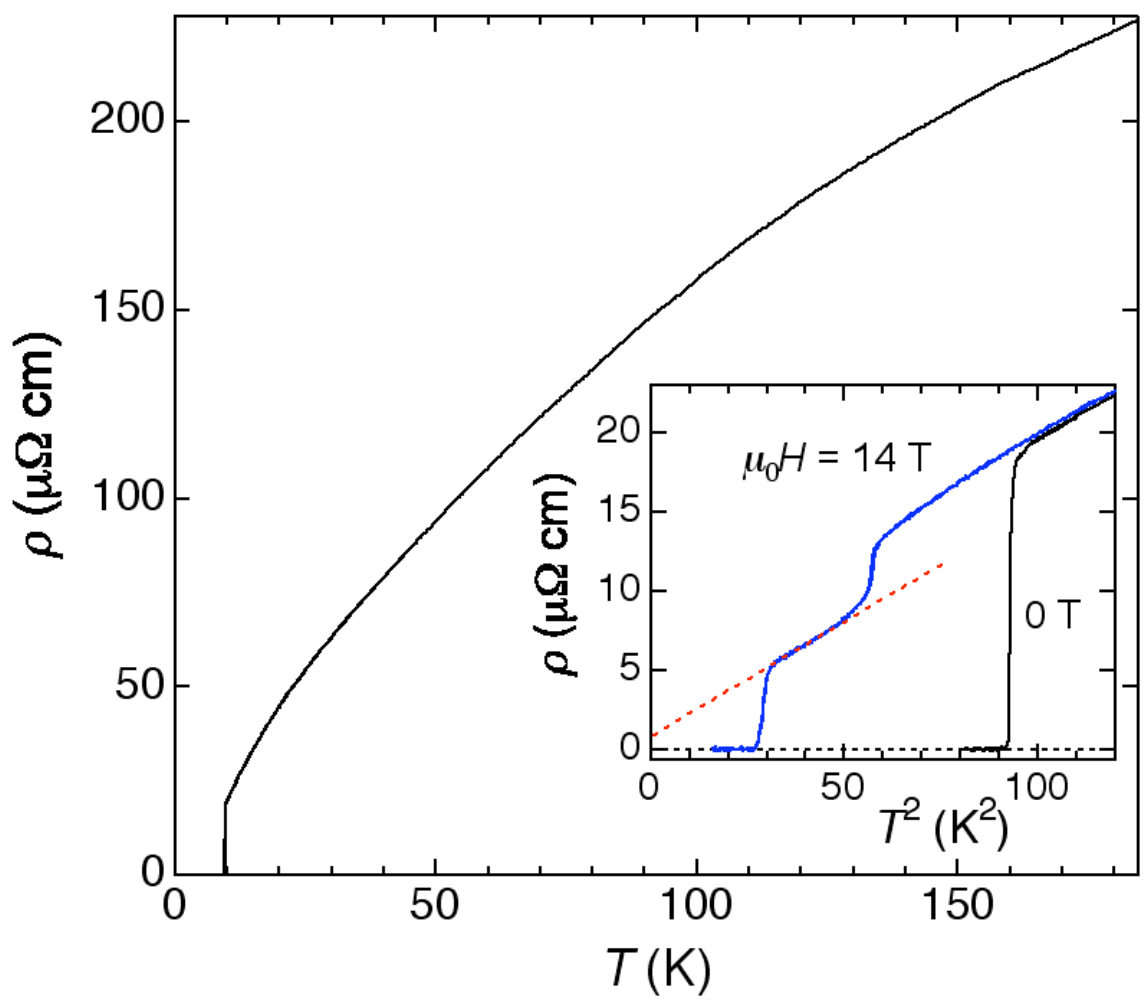

**Figure 4**. (Color online) Temperature dependence of resistivity for $KOs_2O_6$ measured at zero field showing a concave-downward curvature over a wide temperature range. The inset shows two sets of low-temperature data measured at zero field and 14 T as a function of $T^2$.

A marked feature in resistivity is observed in the intermediate field regime, where $T_c$ and $T_p$ are close to each other, as shown in figure 3b. Resistivity curves measured at 63, 65 and 66 kOe exhibit a zero-resistivity transition at $T_{c0}$ and then a sudden increase near $T_p$ to a finite-resistivity value, followed by a gradual decrease again to a zero-resistivity state. Note that the increase at $T_p$ occurs suddenly, within 10 mK. The resistivity decreases in a few steps at higher fields, where $T_c$ is lower than $T_p$. This re-entrant behavior can probably be ascribed to an abrupt change in $H_{c2}$ at $T_p$, as discussed later. Further re-entrant behavior in resistivity measured on the same crystal as used in the present study was found at low magnetic fields below 20 kOe [31], which provides evidence of anomalous flux pinning in this superconductor.

Figure 5 shows magnetization measured on cooling at three fields near the re-entrant transition, together with the corresponding resistivity. The 65-kOe data show that the diamagnetic response due to the Meissner effect does increase smoothly with decreasing temperature, but exhibits a dip near the re-entrant transition in resistivity, indicating that the superconducting volume fraction is partly reduced below $T_p$. Moreover, such a dip in magnetization is also observed for the 60-kOe data, where re-entrant behavior is absent. We observed a similar dip at $T_p$ in magnetization measurements at lower fields down to 10 kOe. Thus, superconductivity is affected by the second phase transition not only near the re-entrant transition, but also in a wide field range.

Figure 6 shows the field dependences of $T_c$, $T_{c0}$ and $T_p$ determined by various measurements in a *H*-*T* phase diagram. All the second anomalies observed in specific heat, resistivity and magnetization measurements are located at $T_p$ = 7.5–7.6 K. Hence, they can all be ascribed to the second phase transition. $H_{c2}$ determined by specific heat measurements increases with decreasing temperature almost linearly. The initial slope is -6.35 T/K, in good agreement with previously reported values [19, 30]. Surprisingly, the linear behavior of $H_{c2}$ changes in slope across the second transition line, which is almost vertical at $T_p$. Moreover, linear extrapolation of $H_{c2}$ above and below $T_p$ causes a distinct gap at $T_p$. Therefore, it can be concluded that $H_{c2}$ changes suddenly across the second transition, dividing the superconducting phase diagram into two regions: the high-temperature phase I has a larger $H_{c2}$ value than the low-temperature phase II. On the other hand, the $H_{irr}$ line determined at $T_{c0}$ in resistivity measurements lies slightly below the $H_{c2}$ line above $T_p$, while they approximately coincide with each other below $T_p$, suggesting that flux pinning is relatively weak in phase I [31]. The $H_{irr}$ line also exhibits a distinct gap at $T_p$, as shown in figure 6b.

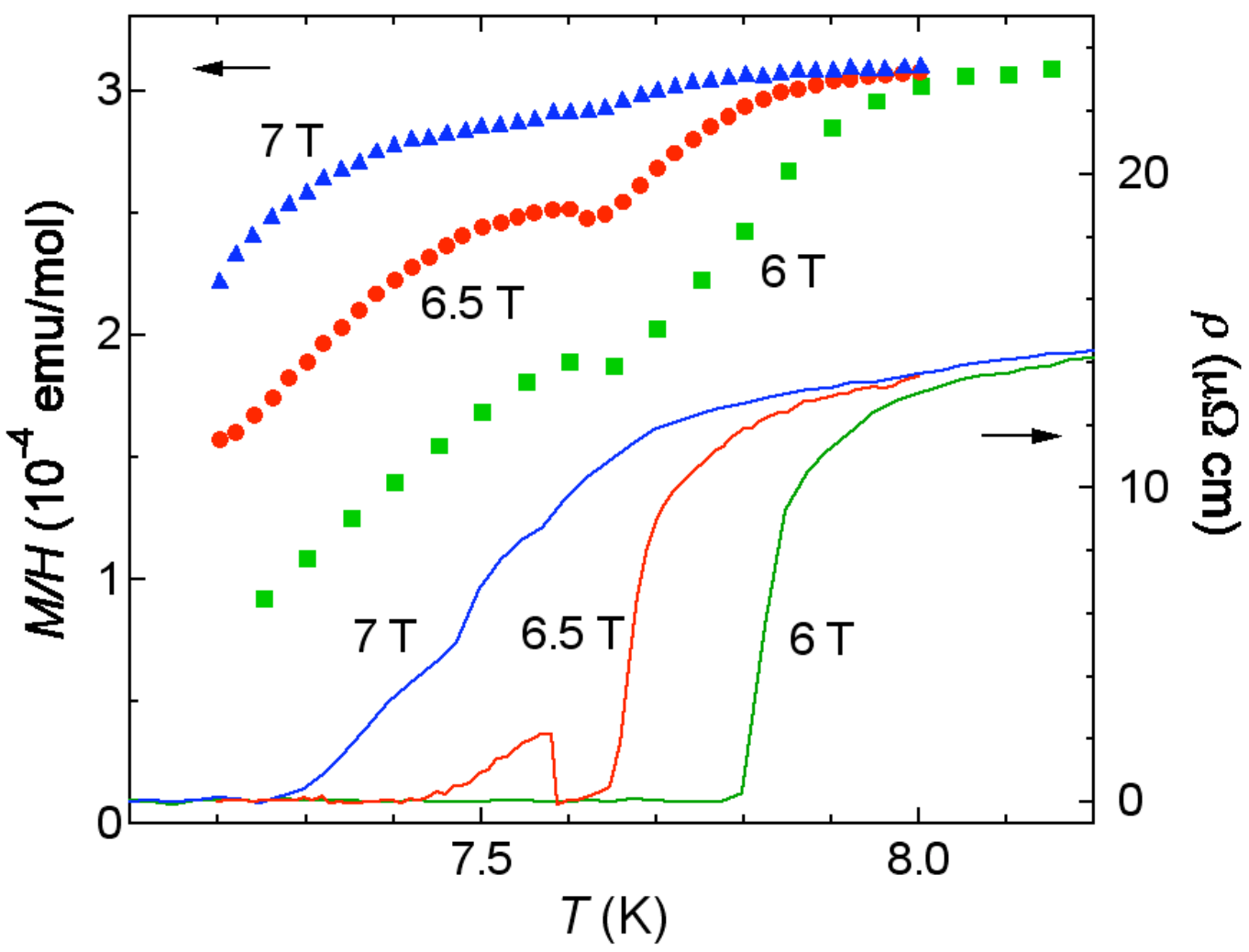

**Figure 5**. (Color online) Magnetization divided by field ($M/H$) measured on cooling in fields of 6.0, 6.5 and 7.0 T applied along the [1-10] direction. Corresponding sets of resistivity data are also shown.

The re-entrant behavior observed in the temperature dependence of resistivity reminiscent of a peak effect in type-II superconductors that is ascribed to enhanced flux pinning near the $H_{c2}$ line. The sudden recovery of resistivity below $T_p$ may be explained if we assume that flux pinning is weaker in low-temperature phase II than in high-temperature phase I. However, this may not be the case, as described above. Alternatively, the re-entrant behavior may be directly related to the change in $H_{c2}$ at $T_p$. As shown in figure 6b, the $H_{c2}$ lines above and below $T_p$ seem to have a fault at $T_p$ in their linear extrapolation. Since we could not determine $H_{c2}$ near $T_p$ in our specific heat measurements because of the existence of a huge peak at $T_p$, there is an ambiguity in the validity of this linear extrapolation of the $H_{c2}$ lines near $T_p$. Nevertheless, it is plausible to assume that $H_{c2}$ varies smoothly in each phase region and changes suddenly at $T_p$ due to the first-order phase transition. Certainly, the Meissner onset temperature varies linearly near $T_p$, as shown in figure 6b. If so, straightforward re-entrant behavior is expected when the sample is cooled in a field near 65 kOe. On cooling at $H = 65$ kOe, for example, bulk superconductivity may set in at $T_c \approx 7.9$ K to reach a zero-resistivity state at $T_{c0} \approx 7.6$ K, and is suppressed at $T_p = 7.57$ K, because the $H_{c2}$ of phase II is lower than 65 kOe just below $T_p$. Then a second zero-resistivity transition takes place at 7.43 K.

The origin of the second phase transition found in $KOs_2O_6$ is now discussed briefly. The possibility of another superconducting transition or magnetic order must be excluded, because of the insensitivity of $T_p$ to magnetic fields and because of the absence of anomalies at $T_p$ in magnetization at high fields up to 140 kOe. Here we assume a structural phase transition that could provide us with reasonable interpretations of the present experimental results. It may be related to the rattling of K ions in an unharmonic potential created by the Os-O cage. From the specific heat data at high temperature, we estimated the characteristic energy of the rattling motion to be less than 40 K assuming the Einstein model, which is unusually small in a solid. Thus, large entropy associated with this rattling freedom can survive down to low temperature, even below $T_c$. It is plausible that this entropy is partially released by the first-order phase transition at $T_p$. The observed entropy change is rather small, 360 mJ $K^{-1}$ $mol^{-1}$, compared with the value of $R\ln 4 \approx 11.5$ J $K^{-1}$ $mol^{-1}$ expected for a case with four potential minima existing in a cage [25]. A preliminary single-crystal X-ray diffraction study at low temperature revealed that the crystal retains its cubic symmetry below $T_p$ and that a structural change associated with the second transition would be subtle, if any [26]. This may be reasonable, because structural couplings between the rattlers and cage must be weak. In any case, experimental efforts to detect a possible structural change at $T_p$ is in progress.

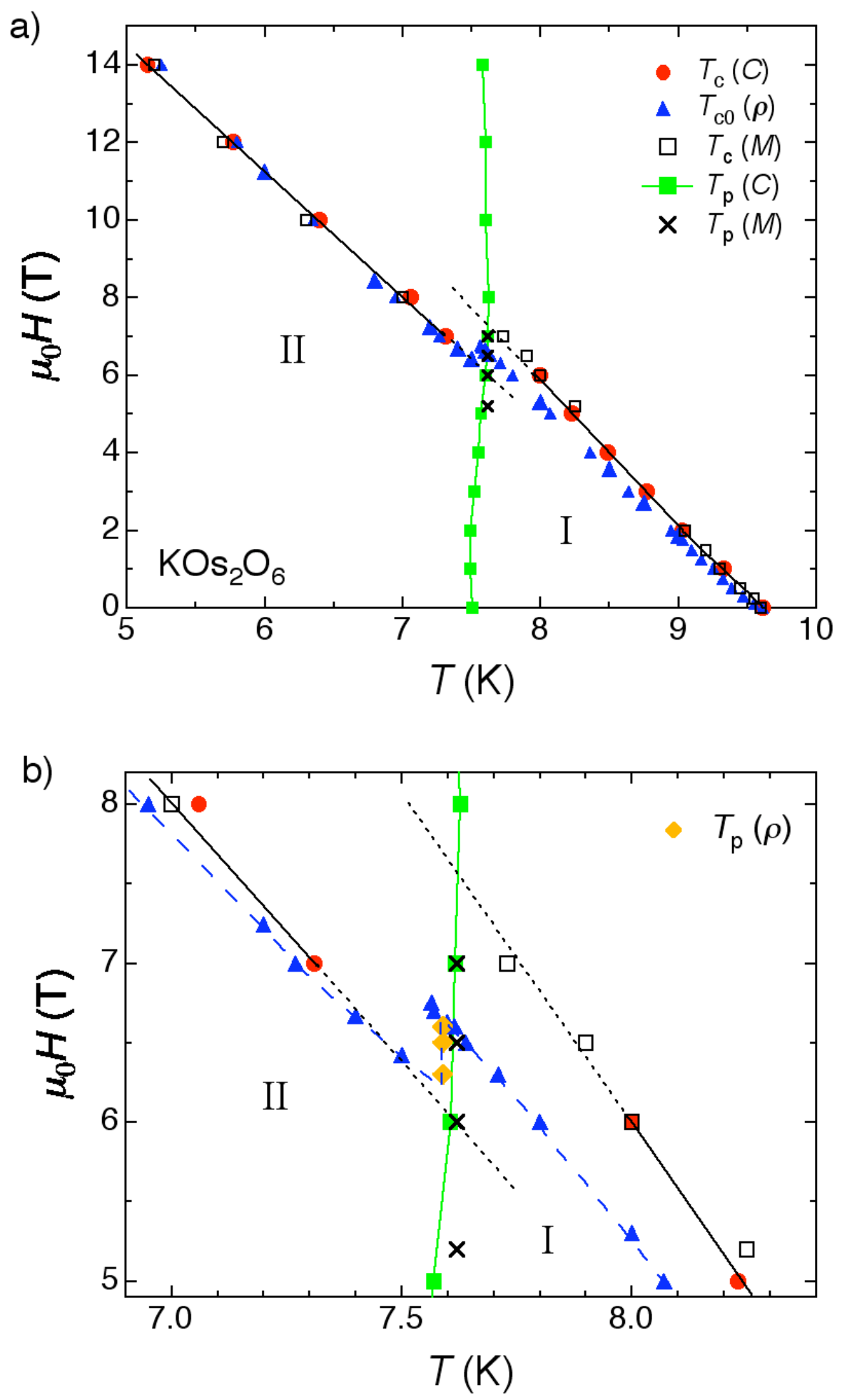

**Figure 6.** (Color online) *H*-*T* phase diagram for $KOs_2O_6$. Circles indicate $H_{c2}$ data determined from specific heat measurements. Triangles indicate $T_{c0}$, defined as the zero-resistivity temperature. Open squares show the Meissner onset temperature, determined from magnetization measurements on cooling. Solid squares represent the peak temperature for specific heat, and crosses represent the position of the dip found in magnetization measurements. Diamonds in (b) represent temperatures at which resistivity exhibited a jump in the re-entrant behavior of resistivity shown in figure 3b. Solid lines on circles represent the $H_{c2}$ curves, and the dotted lines drawn by linear extrapolation. Broken lines on triangles serve as a visual guide.

The observed reduction in resistivity and change from concave-downward to $T^2$ behavior across $T_p$ can be interpreted by assuming an appropriate electron-phonon interaction between the rattling K ions and conduction electrons confined in the Os-O cage. The K ions possess well-defined 1+ charge, because the K 4*s* state is located far above the Fermi energy level and thus is empty according to band structure calculations [25, 32]. Moreover, the electronic structure near the Fermi level is dominated by strongly hybridized Os 5*d* and O 2*p* states. Therefore, a significant interaction is expected between them, because the charge of the rattler must be effectively screened by surrounding conduction electrons on the cage. If electron scattering due to rattling, which may give rise to concave-downward resistivity, is suppressed below $T_p$ and the lifetime of carriers are enhanced, so that the resistivity is reduced. The $T^2$ behavior below $T_p$ may arise from other types of electron-phonon scattering or electron-electron scattering. The large enhancement of the quasiparticle mean free path below $T_p$ was recently reported by Kasahara *et al* in their thermal conductivity measurements [21].

The influence of rattling and its transition on the superconductivity is not clear at present and needs to be clarified in future work. However, it is likely that the low-energy rattling mode plays an important role in enhancing $T_c$, as in A-15 or Chevrel phase compouds [3]. The uniqueness of the present $\beta$-pyrochlore oxide superconductors is the structural duality of the rattlers and the cage, which will provide an interesting model to study the relationship between low-energy phonons and superconductivity in a systematic way.

## 4. Conclusions

We reported experimental evidence to show the existence of a second first-order phase transition in the $\beta$-pyrochlore oxide superconductor $KOs_2O_6$. The influence of this transition on the normal and superconducting properties was described. The source of this structural phase transition is ascribed to the rattling of K cations.

We are grateful to M. Takigawa, Y. Matsuda and W. E. Pickett for enlightening discussions. This research was supported by a Grant-in-Aid for Scientific Research B (16340101) provided by the Ministry of Education, Culture, Sports, Science and Technology, Japan.

## References

[1] Stewart G R 1984 *Rev. Mod. Phys.* **56** 755
[2] Berthier C, Jerome D, Molinie P and Rouxel J 1976 *Solid State Commun* **19** 131
[3] Testardi L R 1975 *Rev. Mod. Phys.* **47** 637
[4] Furuyama N, Kobayashi N and Muto Y 1989 *Phys. Rev. B* **40** 4344
[5] Kresin V Z and Parkhomenko V P 1975 *Sov. Phys. Solid State* **16** 2180
[6] Hanawa M, Muraoka Y, Tayama T, Sakakibara T, Yamaura J and Hiroi Z 2001 *Phys. Rev. Lett.* **87** 187001
[7] Yamaura J and Hiroi Z 2002 *J. Phys. Soc. Jpn.* **71** 2598
[8] Yonezawa S, Muraoka Y and Hiroi Z 2004 *J. Phys. Soc. Jpn.* **73** 1655
[9] Yonezawa S, Muraoka Y, Matsushita Y and Hiroi Z 2004 *J. Phys. Soc. Jpn* **73** 819
[10] Kazakov S M, Zhigadlo N D, Brühwiler M, Batlogg B and Karpinski J 2004 *Supercond. Sci. Technol.* **17** 1169
[11] Brühwiler M, Kazakov S M, Zhigadlo N D, Karpinski J and Batlogg B 2004 *Phys. Rev. B* **70** 020503R
[12] Yonezawa S, Muraoka Y, Matsushita Y and Hiroi Z 2004 *J. Phys.: Condens. Matter* **16** L9

[13] Muramatsu T, Takeshita N, Terakura C, Takagi H, Tokura Y, Yonezawa S, Muraoka Y and Hiroi Z 2005 *Phys. Rev. Lett.* **95** 167004

[14] Koda A, Higemoto W, Ohishi K, Saha S R, Kadono R, Yonezawa S, Muraoka Y and Hiroi Z 2005 *J. Phys. Soc. Jpn.* **74** 1678

[15] Magishi K, Matsumoto S, Kitaoka Y, Ishida K, Asayama K, Uehara M, Nagata T and Akimitsu J 1998 *Phys. Rev. B* **57** 11533

[16] Khasanov R, Eshchenko D G, Karpinski J, Kazakov S M, Zhigadlo N D, Brutsch R, Gavillet D, Castro D D, Shengelaya A, Mattina F L, Maisuradze A, Baines C and Keller H 2004 *Phys. Rev. Lett.* **93** 157004

[17] Khasanov R, Eshchenko D G, Castro D D, Shengelaya A, Mattina F L, Maisuradze A, Baines C, Luetkens H, Karpinski J, Kazakov S M and Keller H 2005 *Phys. Rev. B* **72** 104504

[18] Schneider T, Khasanov R and Keller H 2005 *Phys. Rev. Lett.* **94** 077002

[19] Brühwiler M, Kazakov S M, Karpinski J and Batlogg B 2006 *Phys. Rev. B* **73** 094518

[20] Schuck G, Kazakov S M, Rogacki K, Zhigadlo N D and Karpinski J 2006 *Phys. Rev. B* **73** 144506

[21] Kasahara Y, Shimono Y, Shibauchi T, Matsuda Y, Yonezawa S, Muraoka Y and Hiroi Z 2006 *cond-mat/0603036*

[22] Arai K, Kikuchi J, Kodama K, Takigawa M, Yonezawa S, Muraoka Y and Hiroi Z 2005 *Physica B* **359** 488

[23] Hiroi Z, Yonezawa S, Yamaura J, Muramatsu T and Muraoka Y 2005 *J. Phys. Soc. Jpn.* **74** 1682

[24] Hiroi Z, Yonezawa S, Muramatsu T, Yamaura J and Muraoka Y 2005 *J. Phys. Soc. Jpn.* **74** 1255

[25] Kunes J, Jeong T and Pickett W E 2004 *Phys. Rev. B* **70** 174510

[26] Yamaura J, Yonezawa S, Muraoka Y and Hiroi Z 2005 *J. Solid State Chem.* **179** 336

[27] Yonezawa S, Kakiuchi T, Sawa H, Yamaura J, Muraoka Y, Sakai F and Hiroi Z accepted for *Solid State Sciences*

[28] Nolas G S, Cohn J L, Slack G A and Schujman S B 1998 *Appl. Phys. Lett.* **73** 178

[29] Keppens V, Mandrus D, Sales B C, Chakoumakos B C, Dai P, Coldea R, Maple M B, Gajewski D A, Freeman E J and Bennington S 1998 *Nature* **395** 876

[30] Hiroi Z, Yonezawa S, Yamaura J, Muramatsu T, Matsushita Y and Muraoka Y 2005 *J. Phys. Soc. Jpn.* **74** 3400

[31] Hiroi Z and Yonezawa S 2006 *J. Phys. Soc. Jpn.* **75** 043701

[32] Saniz R, Medvedeva J E, Ye L H, Shishidou T and Freeman A J 2004 *Phys. Rev. B* **70** 100505